\documentclass[a4paper]{article}

\usepackage[british]{babel}

\usepackage[pdftex]{color,graphicx}

\usepackage[T1]{fontenc}
\usepackage[latin9]{inputenc}
\usepackage{url}

\usepackage{amsmath}
\usepackage{amssymb}

\usepackage{hyperref}
\hypersetup{plainpages = false,
              breaklinks=true,
              a4paper=true,
              linktocpage,
              pdfauthor={Niko Brummer},
              pdfstartpage=1, 
              pdfstartview=FitH,  
              pdfkeywords={}}

\title{The BOSARIS Toolkit: Theory, Algorithms and Code for Surviving the New DCF}

\makeatletter
\def\name#1{\gdef\@name{#1\\}}
\makeatother
\author{Niko Br\"ummer and Edward de Villiers\\AGNITIO Research, South Africa}
\date{December 2011}

\newcommand{\effpi}{\tilde\pi}
\newcommand{\Ptar}{P_\text{tar}}

\newcommand{\Pmiss}{P_\text{miss}}
\newcommand{\Pfa}{P_\text{fa}}
\newcommand{\minDCF}{\text{minDCF}}
\newcommand{\TarSet}{\mathcal{T}}
\newcommand{\NonSet}{\mathcal{N}}
\newcommand{\indic}{I}
\newcommand{\Cmiss}{C_\text{miss}}
\newcommand{\Cfa}{C_\text{fa}}
\newcommand{\alphavec}{\boldsymbol{\alpha}}

\newcommand{\funcdef}[3]{#1:#2\mapsto#3}
\newcommand{\R}{\mathbb{R}}

\DeclareMathOperator{\logit}{logit}
\DeclareMathOperator{\probit}{probit}
\DeclareMathOperator{\erf}{erf}

\def\DCF{\text{DCF}}
\def\calR{\mathcal{R}}
\def\calL{\mathcal{L}}
\def\calE{\mathcal{E}}
\def\ND{\mathcal{N}}
\def\minE{\mathcal{E}_\text{min}}
\def\Cllr{C_\text{llr}}

\begin{document}
\maketitle
\begin{abstract}
The change of two orders of magnitude in the \emph{new DCF} of SRE'10,
relative to the old DCF evaluation criterion, posed a difficult challenge
for participants and evaluator alike. Initially, participants were at
a loss as to how to calibrate their systems, while the evaluator
underestimated the required number of evaluation trials.  After the
fact, it is now obvious that both calibration and evaluation require
very large sets of trials. This poses the challenges of (i) how to decide what number of trials is enough, and (ii) how to process such large data sets with reasonable memory and CPU requirements.

After SRE'10, at the BOSARIS 
Workshop, we built solutions to these problems into the freely available BOSARIS 
Toolkit. This paper explains the principles and algorithms behind this
toolkit. The main contributions of the toolkit are:
\begin{enumerate}
\item The \emph{Normalized Bayes Error-Rate Plot}, which analyses likelihood-ratio calibration over a wide range of DCF operating points. These plots also help in judging the adequacy of the sizes of calibration and evaluation databases.
\item Efficient algorithms to compute DCF and minDCF for large score files, over the range of operating points required by these plots.
\item A new score file format, which facilitates working with very
large trial lists.
\item A faster logistic regression optimizer for fusion and calibration.
\item A principled way to define equal error rate, which is of
  practical interest when the absolute error count is small.
\end{enumerate}
\end{abstract}

\section{Introduction}
\label{sec:intro}
The BOSARIS Toolkit provides MATLAB code for calibrating, fusing and evaluating scores from (automatic) binary classifiers. It was developed to provide solutions for automatic speaker recognition, but we envision that much of the code will have wider applicability for other biometric and/or forensics problems, where the calibration of likelihood-ratios is of interest. This document serves as a \emph{user guide}, to explain theory and algorithms and is complementary to the user manual.

The theory behind the toolkit is based on the Ph.D.\ dissertation~\cite{phd:Brummer}, which can be consulted for further details. The core implementation (code) was written by the authors of this document, as part of the ABC: AGNITIO, BUT, CRIM submission for the 2010 NIST Speaker Recognition Evaluation (SRE'10)~\cite{web:sre10}. After the evaluation, at the BOSARIS Workshop,\footnote{See \url{http://speech.fit.vutbr.cz/workshops/bosaris2010}} we collaborated with a wider group of researchers to make these algorithms available in toolkit form.\footnote{Available at: \url{http://sites.google.com/site/bosaristoolkit/}}  

This document is organized in three sections: \textbf{Theory} is the bulk of the document, which explains what the toolkit does and why. \textbf{Algorithms} explains how the toolkit does it. \textbf{Code} gives a high-level summary of the implementation. 

\section{Theory}
This section provides the theoretical framework which is necessary for a good understanding of the BOSARIS Toolkit. For the typical speaker recognition expert, part of this material should be very familiar, while other parts may be new. All readers should nevertheless review the familiar parts, where the terminology for discussing the new material will be established. This section is organized as follows: 
\begin{itemize}
	\item Subsection~\ref{subsec:sampling} discusses the problem of running out of errors and the way this is addressed in the toolkit. 
	\item Subsection~\ref{subsec:bayestheory} reviews Bayes decision theory, while~\ref{subsec:dcf} reviews NIST's DCF criterion for evaluating goodness of decisions. 
	\item Subsection~\ref{subsec:bayes_risk} introduces the idea that we can evaluate system outputs in the form of \emph{likelihood-ratios}, rather than decisions. The key is to let the evaluator make the decisions, at the theoretically optimal Bayes threshold. Subsection~\ref{subsec:bayes_err} develops this idea into practical evaluation criteria.
	\item Subsection~\ref{subsec:roc} discusses perhaps unfamiliar relationships between the familiar evaluation tools, ROC/DET, EER and minDCF. 
	\item Subsection~\ref{subsec:fusion} discusses solutions for fusing and calibrating scores.
\end{itemize} 

\subsection{Sampling effects}
\label{subsec:sampling}
All of the evaluation methods used in this toolkit explicitly or implicitly depend on estimating various error-rates by counting occurrences of those errors in a supervised evaluation database. The error-rates depend not only on the accuracy of the system under evaluation, but also on the \emph{operating point}. We explain operating points in more detail later. What is important here is that no matter what the accuracy of the system under evaluation, or no matter what the size of the evaluation database, there will be operating points where the error-rates become so small that no more errors are observed. More generally, there will be operating points where the numbers of observed errors become so small that the error-rate estimates become unreliable.

There are various frequentist (confidence interval) or Bayesian (credible interval) methods to theoretically quantify the accuracy of such estimates---see for example~\cite{art:Brown_2001} and references therein. The results of any such analysis will depend on various modelling assumptions. 

For the speaker recognition problem, one such analysis, \emph{Doddington's Rule of 30}~\cite{inproc:DoddingtonsRule_98}, is rendered tractable via the assumption of independent Bernoulli trials.\footnote{Are different scores of the same speaker independent? Are miss and false-alarm rates independent?} This rule suggests one needs at least 30 errors to get a probably approximately correct error-rate estimate. In practice, we have found this rule to work well. We get sensible results in both training and test, if we ensure that there are at least 30 misses and at least 30 false-alarms at the operating point of interest. 

\subsubsection{Toolkit solution}
In the BOSARIS Toolkit, we address the problem by flagging on our plots (DET curves as well as normalized Bayes error-rate curves) the points at which the various error-rates drop below 30. It is up to the user of the toolkit to understand that regions on the plot beyond these flags must be treated with caution.

\subsubsection{In SRE'10}
In SRE'10, at the `new DCF' operating point of interest, there was a scarcity of false-alarms, which we addressed by manufacturing many more non-target trials. (This was possible because the number of possible non-target trials grows quadratically with the number of speakers in the available data.) 

We used the rule of thumb that: 
\begin{itemize}
	\item If we want to use a database for calibration/fusion, that database has to be sufficiently large so that the calibrated/fused system makes at least 30 \emph{training} errors of both types, at all operating points of interest.
	\item If we want to use an independent database for testing/evaluation, the same holds. That database has to be sufficiently large so that the system makes at least 30 \emph{test} errors of both types, at all operating points of interest.
\end{itemize}

\subsection{Bayes decision theory}
\label{subsec:bayestheory}
The toolkit is focused on the canonical \emph{speaker detection} problem, where independent decisions must be made for independent trials, based on the output \emph{scores} of an automatic speaker recognition system. In most of this section, we consider the case of making decisions by using the output scores of a single system. We defer fusion of multiple systems to subsection~\ref{subsec:fusion}. 

The input to the toolkit is in the form of \emph{scores} calculated by the automatic system. We assume that for every trial, the system has calculated a scalar detection score and that a decision has to be made based on this score. The recipe for doing so is given by Bayes decision theory~\cite{book:DeGroot_Opt_Stat_1970}.

In the canonical detection problem, there are two alternative hypotheses, called \emph{target} and \emph{non-target}, exactly one of which must be true for every trial. By convention, larger (more positive) scores favour the target hypothesis and smaller (more negative) scores favour the non-target hypothesis. 

For every trial, an \emph{accept/reject} decision is required. We define the \emph{outcome} of a trial as the pair $(\text{hypothesis},\text{decision})$, so that there are four possible  outcomes. Two of these are considered to be \emph{errors}: $\text{miss}=(\text{target},\text{reject})$ and  $\text{false-alarm}=(\text{non-target},\text{accept})$. The other two outcomes are the correct outcomes.

The \emph{consequence} of an outcome is expressed as a \emph{cost function}, which maps outcomes to positive real numbers. Without loss of generality (see~\cite[section 3.4]{phd:Brummer} and~\cite{book:DeGroot_Opt_Stat_1970}), we restrict attention to cost functions which assign zero cost to correct outcomes. This leaves two costs to be specified: $\Cmiss$, the cost of a miss; and $\Cfa$, the cost of a false-alarm.

When given a score, say $s$, the Bayes decision chooses the option, accept or reject, that \emph{minimizes the risk}. That is, we choose to accept if
\begin{align}
P(\text{target}|s,\pi) \Cmiss &\ge P(\text{non-target}|s,\pi) \Cfa
\end{align} 
and to reject otherwise. The two risks being compared are products of costs and posterior probabilities. The posteriors are conditioned not only on $s$, but also on some independent \emph{prior information}, which we represent as:
\begin{align}
\pi &= P(\text{target}) = 1- P(\text{non-target})
\end{align}
We refer to $\pi$ as the \emph{target prior}, or simply as the \emph{prior}. By using Bayes' rule and taking logs, we can rewrite the decision rule as follows:
\begin{align}
\text{accept, if }\ell(s) &\ge \eta, \text{or reject otherwise.}
\end{align} 
where we have defined the \emph{log-likelihood-ratio}:
\begin{align}
\ell(s) &=\log\frac{P(s|\text{target})}{P(s|\text{non-target})}
\end{align} 
the \emph{Bayes decision threshold}:
\begin{align}
\eta = \log\frac{\Cfa}{\Cmiss} - \logit\pi
\end{align} 
and the \emph{prior log odds}:\footnote{The invertible function $\logit(p)=\log\frac{p}{1-p}$ maps probabilities in $[0,1]$ to log odds in $[-\infty,\infty]$.}
\begin{align}
\logit\pi &= \log\frac{\pi}{1-\pi}
\end{align}

We refer to the function $\funcdef{\ell}{\R}{\R}$ as the \emph{calibration mapping}. It maps the score, $s$, to the log-likelihood-ratio, $\ell(s)$. Since log-likelihood-ratios follow the same convention as the scores (larger values favour the target hypothesis), they are also scores. We shall therefore also refer to them as \emph{calibrated scores}. On the other hand, scores are generally not calibrated and cannot do the work of log-likelihood-ratios: when scores are thresholded at the Bayes decision threshold, they usually do not make good decisions. 

The toolkit is concerned with: (i) evaluating the potential ability of the scores, $s$, to make Bayes decisions, even if the calibration mapping, $\ell$, is not available; (ii) creating such mappings, by training on a supervised calibration database; and (iii) evaluating the ability of the calibrated log-likelihood-ratios $\ell(s)$ to make Bayes decisions.

\subsection{DCF: criterion for goodness of hard decisions}
\label{subsec:dcf}
The Bayes decision paradigm leads naturally to a recipe for evaluating the goodness of detection decisions made on a database of supervised trials. In the Speaker Recognition Evaluations (SREs) of 1997 to the present (2010), NIST has required systems under evaluation to submit a hard accept/reject decision, as well as a score, for each trial. The primary evaluation criterion, called DCF (detection cost function), evaluated the goodness of the hard decisions, while secondary criteria (minDCF and DET-curves) evaluated the goodness of the scores.\footnote{DCF as defined here is often referred to as \emph{actual} DCF, to distinguish it from minDCF, which will be defined later.} 

In what follows, we shall always assume that hard decisions, if made by the evaluee, are made by thresholding all scores against a single fixed system-dependent threshold, set by each evaluee. If the evaluee believes the scores to be well-calibrated log-likelihood-ratios, then (s)he may use the Bayes decision threshold $\eta$. Otherwise, the threshold may be tuned by the evaluee to minimize DCF on a supervised calibration database. 

The errors that result from the hard decisions on the supervised evaluation database are summarized as the empirical error-rates: $\Pmiss$, the ratio of misses to target trials; and $\Pfa$, the ratio of false-alarms to non-target trials. The primary evaluation criterion is defined as:
\begin{align}
\DCF &= \pi\Cmiss\Pmiss + (1-\pi)\Cfa\Pfa
\end{align}
It is important to realize that $\pi$ is a \emph{synthetic} parameter, which models the target prior in the domain of application. It does not necessarily reflect the proportion of targets in the evaluation database. 

The DCF parametrization, $\pi,\Cmiss,\Cfa$, can loosely be referred to as the DCF \emph{operating point}. The DCF recipe requires the operating point to be \emph{fixed and known} to the evaluee. Below we show how to relax this requirement.

\subsection{Bayes Risk: criterion for goodness of log-likelihood-ratios}
\label{subsec:bayes_risk}
A small modification~\cite{inproc:brummer_odyssey04} to the DCF evaluation recipe makes it applicable to calibrated log-likelihood-ratios, rather than hard decisions: The evaluee submits log-likelihood-ratios (rather than decisions) and \emph{the evaluator makes the decisions}. The requirement for well-calibratedness is enforced by the fact that the evaluator applies the above-defined \emph{Bayes decision threshold}, $\eta$. 

The error-rates now depend on the evaluator's threshold and we indicate this by the notation $\Pmiss(\eta)$ and $\Pfa(\eta)$. Since the submitted log-likelihood-ratios are also scores, it should be clear that if the evaluator were to sweep $\eta$ from $-\infty$ to $\infty$, then $\Pmiss(\eta),\Pfa(\eta)$ would map out the familiar ROC/DET curve. 

Let $\calL=\ell_1,\ell_2,\cdots,\ell_t,\cdots$ be the log-likelihood-ratios computed by the system under evaluation for every trial, $t$, in the whole supervised evaluation database, so that:
\begin{align}
\Pmiss(\eta) &= \frac{1}{|\TarSet|} \sum_{t\in\TarSet} \indic(\ell_t < \eta), &
\Pfa(\eta) &= \frac{1}{|\NonSet|} \sum_{t\in\NonSet} \indic(\ell_t \ge \eta)
\end{align}
where $\indic$ is the indicator function and $\TarSet$ and $\NonSet$ are the sets of indices belonging to target and non-target trials.

The resulting evaluation criterion, the \emph{empirical Bayes risk}, is given by:
\begin{align}
\begin{split}
\calR(\calL|\pi,\Cmiss,\Cfa) &= \pi\Cmiss\Pmiss(\eta) + (1-\pi)\Cfa\Pfa(\eta) \\
\text{where }\eta &= \log\frac{\Cfa}{\Cmiss}-\logit\pi
\end{split}
\end{align}
If the evaluator always applies a fixed, known DCF parametrization, $\pi,\Cmiss,\Cfa$, then nothing essential has changed. For a `calibrated' log-likelihood-ratio, the evaluee could just submit $\ell_t = s_t-\gamma+\eta$, where $s_t$ is his original uncalibrated score and $s_t\ge \gamma$ is his original decision rule. In this case $\mathcal{R}$ would be numerically equal to DCF. 

But, if the evaluator sweeps $\eta$ over a range of values, then everything changes. Now mere shifting will not adequately calibrate the scores. Now scaling as well as finer details of the calibration mapping also matter. (After taking care of a few more details below, we will demonstrate this experimentally.)

The empirical Bayes risk as evaluation criterion for log-likelihood-ratios is discussed in detail in~\cite{art:brummer_csl_2006,incoll:VanLeeuwen_Eval_2007,phd:Brummer}. It can be interpreted as: 
\begin{itemize}
	\item A \emph{proper scoring rule}, which encourages both good discrimination (i.e.\ a good DET-curve) as well as good probabilistic calibration (in the sense of~\cite{art:DeGroot_Forecast_1983}). See for example~\cite{book:Jaynes_PTTLOS_2003}, Chapter 13, the section entitled `The honest weatherman', for an insightful explanation.
	\item \emph{Generalized cross-entropy}~\cite{techrep:UCL_Dawid_139} between the evaluator's perfect empirical posterior given by the labels and the posterior $P(\text{target}|s,\pi)$ of the evaluee. This information-theoretical analysis provides useful inequalities to understand the essential properties of this evaluation criterion~\cite[Chapter 2]{phd:Brummer}.
\end{itemize}

\subsubsection{The default system}
Define the \emph{default system}, which always outputs log-likelihood-ratio of zero, so that $\calL_0=0,0,\cdots$ for every trial. Notice that the posterior of the default system is the same as the prior: $P(\text{target}|\ell_t=0,\pi)=\pi$. Making Bayes decisions with the default system is the same as making decisions with the prior alone.

It is easy to show~\cite[Chapter 2]{phd:Brummer} that if the likelihood-ratios of a system, $\calL$, are sufficiently well calibrated, then $\calR(\calL|\pi,\Cmiss,\Cfa)\le\calR(\calL_0|\pi,\Cmiss,\Cfa)$, for every operating point $\pi,\Cmiss,\Cfa$. A system that fails this test at some operating point can be said to be \emph{badly calibrated} at that operating point. At such operating points, on average, better Bayes decisions are obtained by \emph{not} using the system.

\subsubsection{Simplifying risk to error-rate}
As shown above, a system that outputs well-calibrated likelihood-ratios can be expected to make useful (better than default) Bayes decisions at every operating point. It therefore seems reasonable to expect of an evaluation procedure to test calibration over a wide range of such operating points. The problem is that the Bayes risk, as we have defined it, is parametrized by three independent parameters, $\pi,\Cmiss,\Cfa$. How can we design our evaluation recipe to take account of all operating points in this \emph{three-dimensional} space?

This problem is solved by realizing that all these operating points can be represented by an equivalent one-dimensional range of operating points, which is much easier to cover with an evaluation recipe. We show how this is done.

Define the \emph{effective prior} as:
\begin{align}
\tilde\pi &= \frac{\pi\Cmiss}{\pi\Cmiss+(1-\pi)\Cfa}
\end{align}
and now parametrize the Bayes risk with $\tilde\pi$ and $\tilde{C}_\text{miss}=\tilde{C}_\text{fa}=1$. This reparametrization leaves the Bayes decision threshold, $\eta$, \emph{unchanged}:
\begin{align}
\eta &= -\logit\tilde\pi = \log\frac{\Cfa}{\Cmiss}-\logit\pi
\end{align}
and the evaluation criterion, $\calR$ is merely \emph{scaled}:
\begin{align}
\calR(\calL|\tilde\pi,1,1) &= \frac{1}{\pi\Cmiss+(1-\pi)\Cfa} \calR(\calL|\pi,\Cmiss,\Cfa)
\end{align}
where the scaling factor is positive and is not a function of $\calL$ or of the error-rates. This means that if we are comparing the \emph{relative} benefits of two systems, say $\calL_1$ and $\calL_2$, then:
\begin{align*}
\calR(\calL_1|\tilde\pi,1,1) &\le \calR(\calL_2|\tilde\pi,1,1)
\;\;\Leftrightarrow\;\;
\calR(\calL_1|\pi,\Cmiss,\Cfa) \le \calR(\calL_2|\pi,\Cmiss,\Cfa)
\end{align*}
from which we conclude that the two criteria are \emph{equivalent} for evaluation purposes.\footnote{This equivalence still holds if we allow more general cost functions, which can have negative costs (i.e.\ rewards) for correct decisions. In this case, the relationship between the criteria is affine, rather than linear.~\cite{book:DeGroot_Opt_Stat_1970}}

\subsection{Empirical Bayes error-rate: a practical evaluation recipe}
\label{subsec:bayes_err}
We now define our final evaluation criterion for evaluating the goodness of log-likelihood-ratios. The \emph{empirical Bayes error-rate} is $\calE(\calL|\tilde\pi) = \calR(\calL|\tilde\pi,1,1)$, so that:
\begin{align}
\label{eq:def_ber}
\calE(\calL|\tilde\pi) &= \tilde\pi\Pmiss(-\logit\tilde\pi) + (1-\tilde\pi)\Pfa(-\logit\tilde\pi)
\end{align}
This criterion is parametrized by the \emph{single, scalar} parameter, $\tilde\pi$, or equivalently by the Bayes decision threshold, $-\logit\tilde\pi$. Again, we refer to this parameter as the \emph{operating point}.

The \emph{old operating point} defined by NIST for the SREs between 1997 and 2008 was at $\effpi\approx0.092$, while the \emph{new operating point} of 2010 was at $\effpi=0.001$.

In this toolkit, we are interested in evaluation that spans operating points. By having confined the operating point to one dimension, this becomes do-able. By sweeping over the threshold, this criterion exercises the decision-making ability of log-likelihood-ratios in a similar way that the ROC/DET-curve exercises the potential decision-making ability of uncalibrated scores. In subsections below, we shall discuss two ways of sweeping the operating point: one is an integral, the other a plot.

\subsubsection{The default system: reference for bad calibration}
We provide two references which can be compared to $\calE(\calL|\effpi)$ to judge calibration of $\calL$. The first, discussed here, is the upper boundary where calibration fails. The other (the familiar minDCF), discussed in the next subsection, is an ideal lower bound, where calibration is optimal. 

The default system, $\calL_0$, provides the reference error-rate:
\begin{align}
\calE(\calL_0|\tilde\pi) &= \min(\tilde\pi,1-\tilde\pi)
\end{align}
As mentioned above, a system $\calL$, for which $\calE(\calL|\tilde\pi)>\calE(\calL_0|\tilde\pi)$, is said to be badly calibrated at the operating point $\tilde\pi$, because then it would be better not to use the system.

\subsubsection{minDCF: reference for ideal calibration}
\def\minDCF{\text{minDCF}}
NIST's minDCF is obtained by allowing the \emph{evaluator, who has access to the true class labels}, to choose an optimal threshold at every operating point:
\begin{align}
\minDCF(\calL|\pi,\Cmiss,\Cfa) &= \min_{-\infty\le\gamma\le\infty} \pi\Cmiss\Pmiss(\gamma) + (1-\pi)\Cfa\Pfa(\gamma)
\end{align}
Here we are interested in the specialization of minDCF, where the costs are unity. In analogy with $\calE$, we denote it $\minE$:
\begin{align}
\label{eq:min_ber}
\minE(\calL|\tilde\pi) &= \minDCF(\calL|\effpi,1,1)
\end{align}
Note:
\begin{align}
\calE(\calL|\tilde\pi) \ge \minE(\calL|\tilde\pi) \le \calE(\calL_0|\tilde\pi)
\end{align}
Like minDCF, $\minE$ is a secondary evaluation criterion, which fulfils two functions:
\begin{itemize}
	\item It provides an ideal reference value for judging calibration. If $\calE$ and $\minE$ are close, then the system can be said to be very well calibrated.
	\item In the earlier stages of the development of a speaker recognition algorithm, one is typically not interested in calibration, but just in the \emph{potential} to make good decisions at some operating point.  $\minE$ provides a calibration-insensitive criterion, which can be evaluated over a range of different operating points. 
\end{itemize}

\subsubsection{Cllr: scalar summary of goodness of log-likelihood-ratios}
The BOSARIS Toolkit provides two ways to sweep the operating point: one \emph{integrates} out the operating point to give a scalar, summary criterion; and the other \emph{plots} the error-rate as a function of the operating point. We discuss the integral here and the plot in the next subsection.

We can define the calibration-sensitive, scalar summary criterion of the goodness of log-likelihood-ratios, known as $\Cllr$, by integrating out the operating point~\cite{art:brummer_csl_2006}:
\begin{align}
\begin{split}
\Cllr(\calL) &= k\int_{-\infty}^{\infty} \calE(\calL|\logit^{-1}x) \,dx \\
&= \frac{0.5}{|\TarSet|}\sum_{t\in\TarSet} \log_2(1+e^{-\ell_t}) + \frac{0.5}{|\NonSet|}\sum_{t\in\NonSet} \log_2(1+e^{\ell_t})
\end{split}
\end{align}
where $k>0$ is an unimportant scale factor and $\logit^{-1}x=(1+e^{-x})^{-1}$ is the inverse\footnote{$\logit^{-1}$ is also known as the \emph{logistic sigmoid}.} of the logit function. 

This criterion is further discussed in~\cite{phd:Brummer,art:brummer_csl_2006,incoll:VanLeeuwen_Eval_2007,phd:Ramos}. It can be interpreted as a strictly proper scoring rule, empirical cross-entropy, negative log-likelihood and as optimization objective for logistic regression. 

\subsubsection{Normalized Bayes-error-rate plots}
\label{sec:bayes_plots}
To plot $\calE(\calL|\tilde\pi)$ as a function of the operating point, it is helpful to transform both the horizontal and vertical axes. 

Using $\tilde\pi\in[0,1]$ as the horizontal axis would compress interesting parts of the graph against the sides of the interval. We therefore use $\logit\tilde\pi$ on the horizontal axis instead. This axis now becomes infinite in both directions and we plot only a suitable interval, near the origin, $\logit0.5=0$. Plotting an interval that is too wide is meaningless anyway, because in those regions the prior becomes so close to 0 or 1 that either the miss or the false-alarm counts drop to zero.

The vertical axis is non-linearly amplified by normalizing with $\calE(\calL_0|\tilde\pi)=\min(\tilde\pi,1-\tilde\pi)$. If this were not done, low error-rates would compress all the interesting action against the bottom of the plot. 

\begin{figure}[!htb] 
\centerline{
  \includegraphics[trim = 100 200 100 200 , 
                   width=0.8\textwidth]
                   {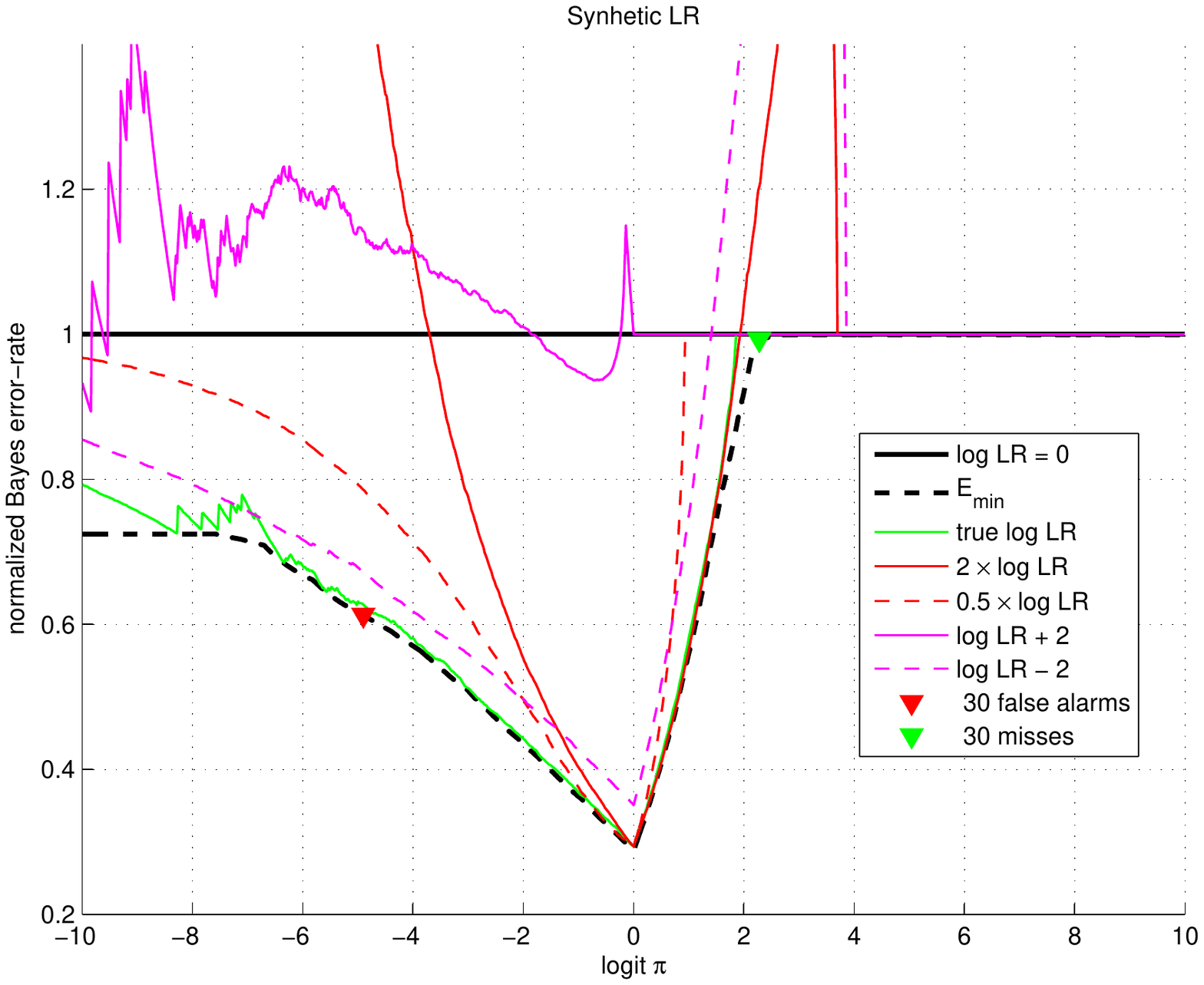}
}
\caption[]{Normalized Bayes error-rate plot for a synthetic system with Gaussian scores: targets $\sim\ND(\mu=3,\sigma=2)$ and non-targets $\sim\ND(0,1)$. The true likelihood-ratio is compared against deliberate additive and multiplicative miscalibrations.}
\label{fig:synthDCF}
\end{figure}

The \emph{normalized Bayes-error-rate plot} can be described as a plot of $(x,y)$ such that:
\begin{align}
y &= \frac{\calE(\calL|\logit^{-1}x)}{\calE(\calL_0|\logit^{-1}x)}
\end{align} 
Figure~\ref{fig:synthDCF} gives an example, using synthetic Gaussian scores to compare the true log-likelihood-ratio against some deliberately miscalibrated versions. This plot demonstrates:
\begin{itemize}
	\item The deliberately miscalibrated `systems' have worse error-rates than the (green) `true LR' system, almost everywhere.
	\item The only region where the green system does worse than the miscalibrated dashed magenta is due to small sample effects. This is to the left of the red triangle, where the number of false-alarms becomes very low. The red and green triangles indicate the points were false-alarms and misses become scarce (less than 30) and therefore indicate the boundaries were small-sample effects may become a problem for meaningful evaluation. The safe region is between the two triangles. The error-rates that determine the horizontal positions of these triangles are obtained from the dashed black curve, where the evaluator has optimized the thresholds. 
	\item The dashed black curve is $\minE(\calL|\tilde\pi)$. Between the triangles, it coincides closely to the theoretically optimal `true LR' green curve. In real cases, we are \emph{not} given a true probability model that generated the data, so that $\minE$ forms a useful practical reference for judging calibration.
	\item The solid black line at $y=1$ represents the default performance of $\calE(\calL_0|\tilde\pi)$. In places, the miscalibrated systems do worse than this reference. The only one which does not is the underoptimistic $0.5\times \log\text{LR}$.
\end{itemize}
(The reason why the dashed and solid black lines meet just to the right of $+2$ for this dataset is that the Gaussian log-likelihood-ratio as a function of the score is a parabola, with a minimum just below $-2$. The system never outputs log-likelihood-ratios with smaller values, so that in the far right of the plot, all decisions are identical to those made by the default system (i.e.\ accept).)

\begin{figure}[!htb] 
\centerline{
  \includegraphics[trim = 100 200 100 200 , 
                   width=0.8\textwidth]
                   {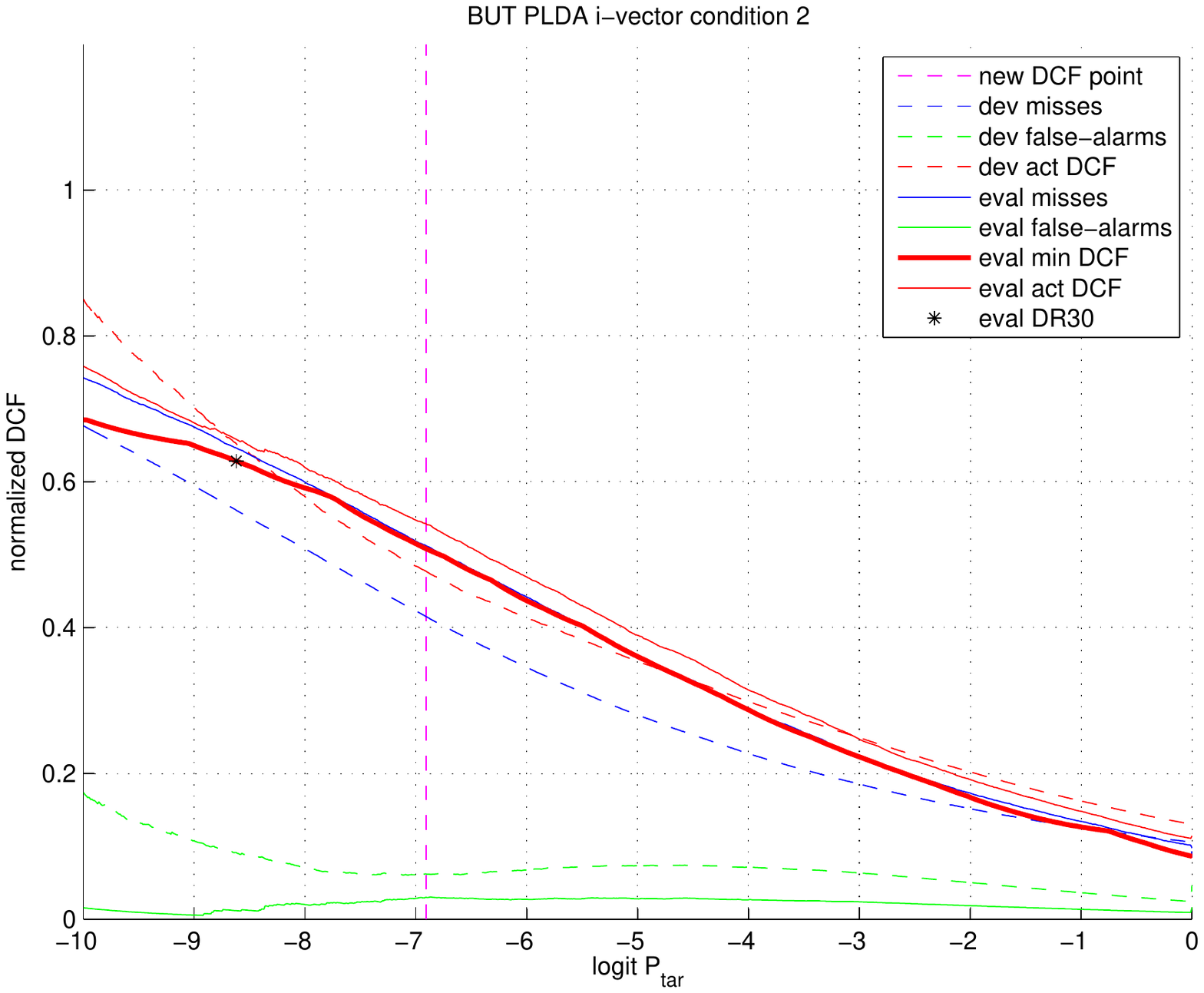}
}
\caption[Normalized Bayes error-rate: good calibration]{Normalized Bayes error-rate plot for an SRE 2010 speaker detector with \emph{good} calibration. Here \emph{eval} denotes the evaluation database and \emph{dev} the development database. $P_\text{tar}=\effpi$, while  act~DCF and min~DCF refer to $\calE$ and $\minE$. $\Pmiss$ and normalized $\Pfa$ are shown separately. DR30 refers to the point to the left of which there are fewer than 30 false-alarms. The vertical magenta dashed line represents the \emph{new operating point} at $\effpi=0.001$.}
\label{fig:lukas}
\end{figure}

\begin{figure}[!htb] 
\centerline{
  \includegraphics[trim = 100 200 100 200 , 
                   width=0.8\textwidth]
                   {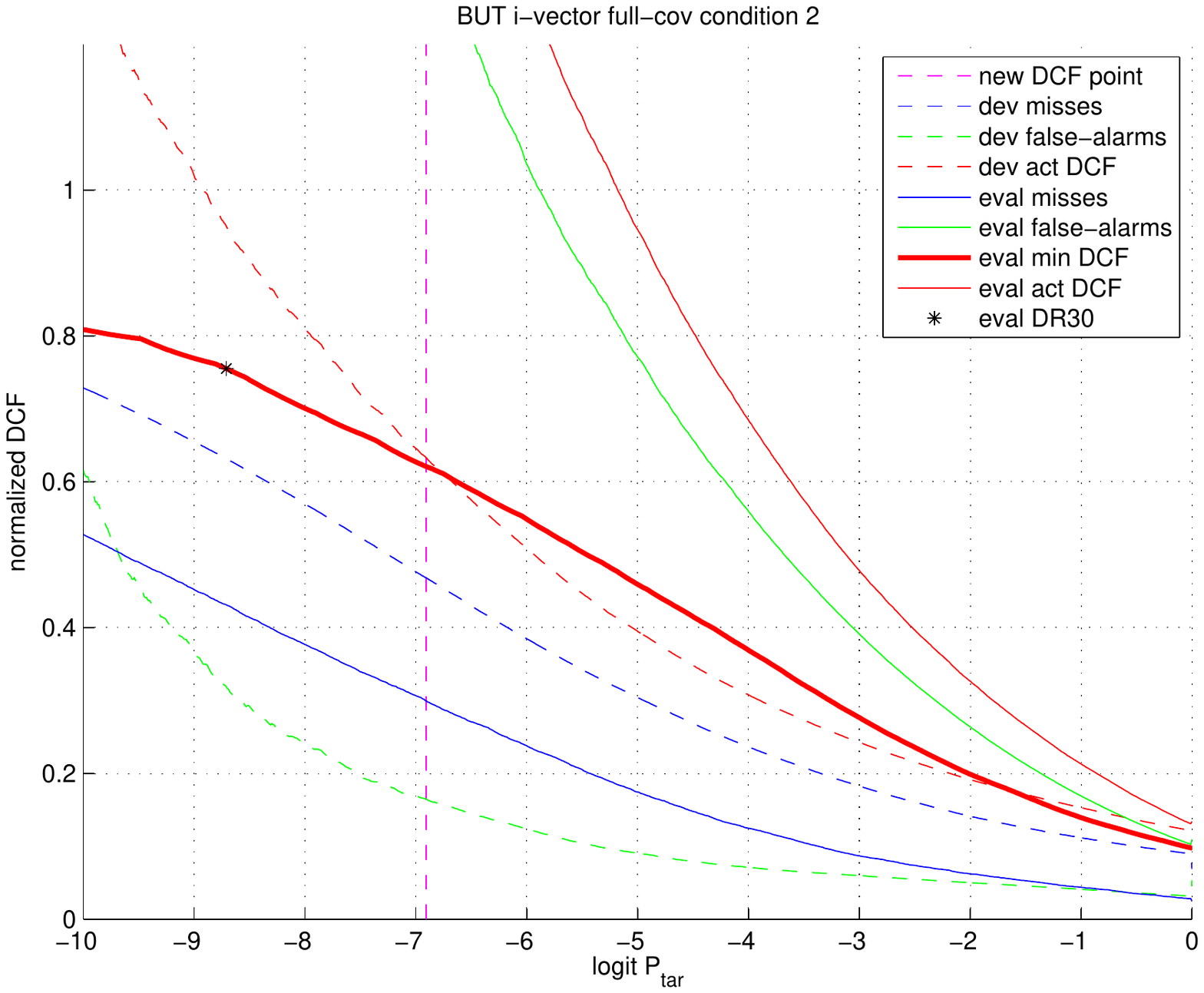}
}
\caption[Normalized Bayes error-rate: bad calibration]{Normalized Bayes error-rate plot for an SRE 2010 speaker detector with \emph{bad} calibration. See caption of figure~\ref{fig:lukas} for details.}
\label{fig:ondra}
\end{figure}

Figures~\ref{fig:lukas} and~\ref{fig:ondra} show further examples of normalized Bayes error-rate plots, but now for real speaker recognition scores of systems submitted to SRE'10.

The plots show curves for tests on two databases: \emph{dev} is the database used to train the calibration (SRE2008 eval database in this case) and \emph{eval} is the evaluation database (SRE2010).  The Bayes error-rate for the dev database is shown in dashed red and that for the eval database in solid red.  The minimum Bayes error-rate (thick red) is only shown for the eval database.  The toolkit can also plot the contributions of the misses and false alarms to both the minimum Bayes error-rate and actual Bayes error-rate.  In the example plots, only the contributions to the actual Bayes error-rate are shown (misses in blue, false alarms in green).  The new (SRE'10) operating point is shown on the plots by the vertical dashed magenta line at $-6.91$.

In the region of interest, $x<0$, which we plot in these figures, the vertical axis (normalized error-rate) is:
\begin{align}
y &= \frac{\calE(\calL|\effpi)}{\min(\effpi,1-\effpi)} \\
&= \frac{\effpi\Pmiss(\eta)+(1-\effpi)\Pfa(\eta)}{\effpi} \\
&= \Pmiss(\eta) + \exp(-\logit\effpi)\Pfa(\eta) \\
&= \Pmiss(-\logit^{-1}x) + \exp(-x)\Pfa(-\logit^{-1}x) 
\end{align}
The exponential amplification of false-alarms induced by this normalization explains the shape of the curves for regions of bad calibration. Some form of amplifying normalization is needed to make the effects of calibration visible in regions of low error-rate. This normalization is the main difference between these curves and APE-curves~\cite{art:brummer_csl_2006}. The normalized Bayes error-rate plot is able to display a wider range of operating points than the APE-curve. 

The points in the plot marked with asterisks (we used triangles in the first plot), labelled DR30 refer to Doddington's Rule of 30~\cite{inproc:DoddingtonsRule_98}. This rule suggests you need at least 30 false-alarms and at least 30 misses for meaningful evaluation.  The toolkit can plot both the DR30 point for the misses (to the right of which the absolute number of misses drops below 30) and the one for the false alarms (to the left of which the absolute number of false-alarms drops below 30). These points are on the $\minE$ curve, because we use the false-alarm count and miss count that result from the evaluator's optimized threshold. 

\subsection{ROC/DET and related criteria for goodness of scores}
\label{subsec:roc}
This subsection deals with ROC/DET curves and associated summaries such as EER and minDCF, all of which can be applied for calibration-insensitive evaluation of the goodness of uncalibrated scores. This is useful for the earlier stages of algorithm development, when calibration is not of immediate interest.

We assume the reader is familiar with the ROC (receiver operating characteristic)~\cite{art:Fawcett_PatRecLet_ROC_2006}. In this section we concentrate on perhaps unfamiliar relationships that exist between the ROC, minDCF and EER. In summary: the ROC spans operating points by plotting error-rates as a function of the threshold; minDCF samples the ROC at a fixed operating point; EER summarizes the span of operating points by \emph{maximizing} over minDCF as a function of the operating point. The \emph{ROC convex hull} is central to this analysis and also provides the key to efficient minDCF and EER calculation.

In our discussion below, we use the term ROC, but (unless otherwise noted) everything applies also to DET-curves~\cite{inproc:Martin_EUROSPEECH_1997}. For ROC, we assume the speaker-recognition convention where $x=\Pfa$ is on the horizontal axis and $y=\Pmiss$ on the vertical axis.\footnote{In other fields, the vertical axis is $1-\Pmiss$.} The DET-curve differs from the ROC by axis warping:\footnote{The probit function maps $[0,1]$ to $[-\infty,\infty]$ in a very similar way to the logit function: $\probit(p)=\sqrt2\erf^{-1}(2p-1)$.} $x=\probit(\Pfa)$ and $y=\probit(\Pmiss)$.  

There are some aspects of the ubiquitous ROC/DET that seem to be misunderstood by many of its users. Here we highlight the following:
\begin{itemize}
	\item The ROC is an \emph{optimistic} view of the decision-making ability of scores, because calibration is not tested. If Bayes risk is minimized (i.e.\ minDCF) at a particular operating point `on the ROC curve', then the calibration problem remains of how to choose a threshold that will place the actual performance at this operating point. This actual performance is usually worse (and cannot be better) than minDCF.  
	\item The empirical ROC is not a continuous curve. It is a collection of discrete points in $(\Pfa,\Pmiss)$ space, where every point corresponds to a decision threshold between adjacent scores. If the points are connected with line segments\footnote{assuming no two scores coincide} then those segments are either vertical or horizontal, corresponding to target and non-target scores. We shall refer to this plot as the \emph{steppy ROC}.
	\item minDCF operating points do not live exactly on the steppy ROC. They live on the \emph{ROCCH curve}: the lower left boundary of the convex hull around the discrete points of the ROC. 
	\item Although the EER is fixed at $\Pmiss=\Pfa$, it nevertheless forms a summary of the whole curve: it is a tight upper bound of the decision making ability over all operating points. Using EER as optimization objective is a good idea, because forcing the tight upper bound down, forces the whole curve down. This can be generalized to any other point on the ROCCH curve by fixing the ratio $\frac{\Pmiss}{\Pfa}$.
\end{itemize}
We elaborate on the last two points below.

\subsubsection{The ROCCH is where minDCF lives}
\label{sec:rocch}
\def\pmiss{p_\text{miss}}
\def\pfa{p_\text{fa}}
\def\alphavec{\boldsymbol{\alpha}}
\def\Vset{\mathcal{V}_{ch}}
Let there be $n$ points, $[\pfa(i),\pmiss(i)]$ in the empirical ROC. A point in $\R^2$ is in the \emph{convex hull} of the ROC, if and only if it is a two-dimensional interpolation between all of the ROC points. That is, a point
\begin{align}
[x,y] = \sum_{i=1}^n \alpha_i[\pfa(i),\pmiss(i)]
\end{align}
is in the convex hull if and only if all $\alpha_i\ge0$ and $\sum_{i=1}^n\alpha_i=1$. 

We already know that minDCF can be expressed either as a continuous minimization over the threshold ($\gamma)$, or as a discrete minimization over the ROC points. But it can also be expressed~\cite{art:provost_mach_learn_2001,phd:Brummer} as a continuous minimization over the convex hull, or as a discrete minimization over the set of vertices, $\Vset$, of the convex hull:
\begin{align}
\label{eq:minDCF}
\begin{split}
\minDCF(\pi,\Cmiss,\Cfa) &= \min_{\gamma} \pi\Cmiss\Pmiss(\gamma)+(1-\pi)\Cfa\Pfa(\gamma) \\
&= \min_{i=1}^n \pi\Cmiss\pmiss(i)+(1-\pi)\Cfa\pfa(i) \\
&= \min_{\alphavec} \sum_{i=1}^n \alpha_i\bigl(\pi\Cmiss\pmiss(i)+(1-\pi)\Cfa\pfa(i)\bigr) \\
&= \min_{i\in\Vset} \pi\Cmiss\pmiss(i)+(1-\pi)\Cfa\pfa(i) 
\end{split}
\end{align}
where $\alphavec=[\alpha_1,\ldots,\alpha_n]$ is subject to the above-mentioned convexity constraint. This means that although parts of the convex hull seem more optimistic than the steppy ROC, these parts do not give lower minDCF, no matter what the operating point.  
 
The DCF minima live on the lower left boundary of the convex hull, which forms a continuous, piecewise linear, convex curve between the points $(0,1)$ and $(1,0)$. We shall refer to this curve as the \emph{ROCCH curve}. 

\begin{figure}[!ht] 
\centerline{
  \includegraphics[trim = 100 200 100 200 , 
                   width=0.8\textwidth]
                   {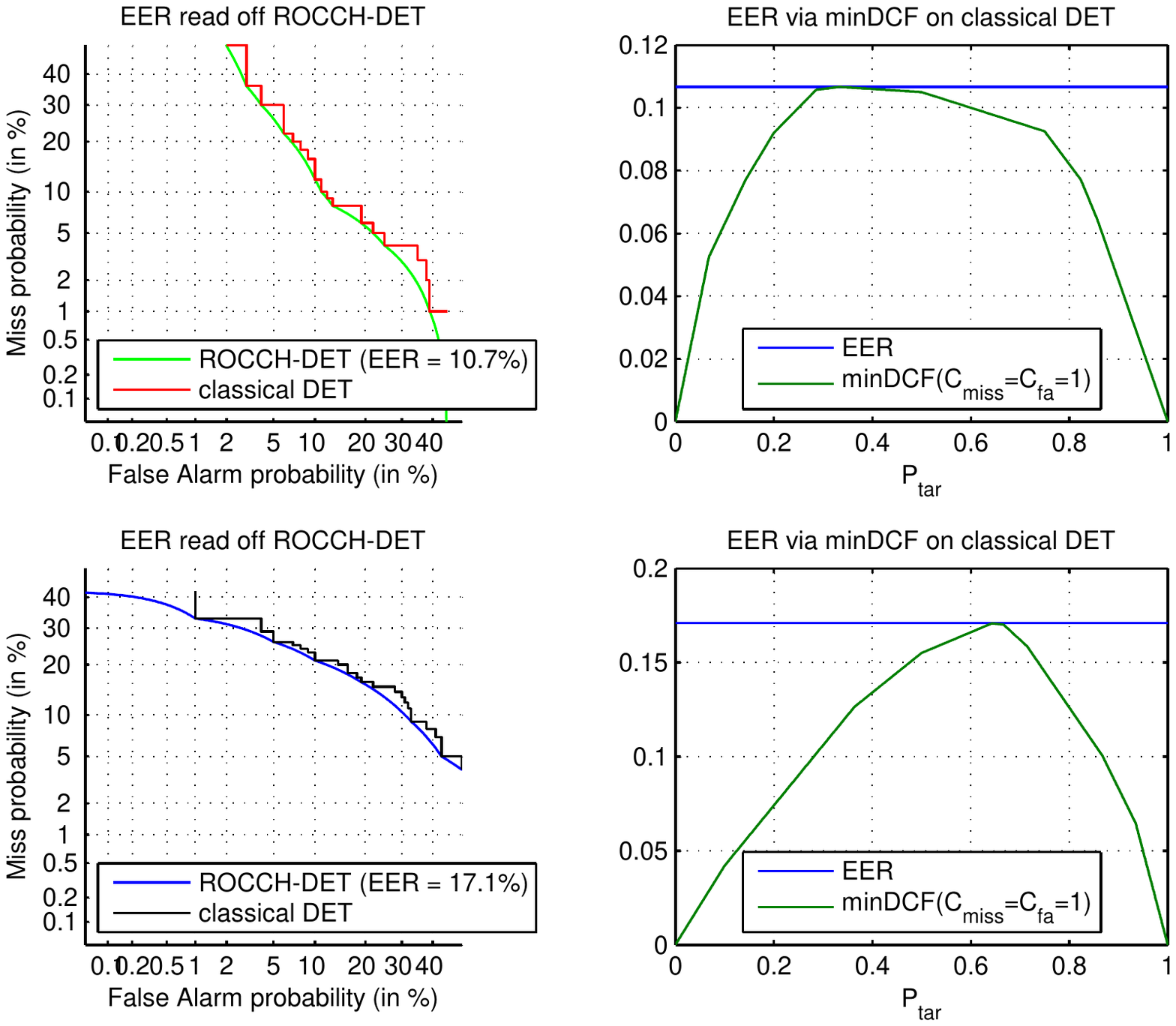}
}
\caption[ROCCH DET curves]{Two examples of ROCCH-DET vs classical steppy DET\@. The equality of ROCCH-EER and max minDCF is demonstrated. (Here $\effpi=\Ptar$).}
\label{fig:rocchdet}
\end{figure}

The BOSARIS Toolkit provides the functionality to compute the ROCCH curve, as well as the associated DET-curve obtained by applying the non-linear (probit) mapping to the axes.\footnote{The convexity does not hold when these curves are translated to DET space.} Figure~\ref{fig:rocchdet} shows two examples. For further examples, see~\cite[Chapter 7]{phd:Brummer}, or~\cite{inproc:Aversano_Evalita_2009}, or try to plot some of your own, using the toolkit.

The ROCCH vertex set, $\Vset$, is typically much, much smaller than the empirical ROC. Since the convex hull can be computed efficiently (see the PAV algorithm below), and since it is valid for all operating points, this is the key to efficient minDCF computations for large score sets, over a large range of operating points. 

\subsubsection{EER as upper bound}
The EER (equal-error-rate) is usually defined as the `point on the ROC', where $\Pmiss=\Pfa$. For the empirical ROC, in general, no point exactly satisfies this equality, but it can be satisfied by interpolation. If we choose to interpolate between \emph{all} points in the ROC, we again find ourselves on the ROCCH curve. We denote the point on the ROCCH curve where $\Pmiss=\Pfa$ as the ROCCH-EER. We propose to use the ROCCH-EER as a well-defined, practical version of the EER and this functionality is provided as such by the toolkit. 

The ROCCH-EER has the following interesting property~\cite{phd:Brummer}:
\begin{align}
\begin{split}
\text{ROCCH-EER} &= \max_{\effpi} \min_{-\infty\le\gamma\le\infty} \effpi\Pmiss(\gamma)  + (1-\effpi)\Pfa(\gamma) \\
&= \max_{\effpi} \minDCF(\effpi,1,1)
\end{split}
\end{align}
Figure~\ref{fig:rocchdet} demonstrates this. 

ROCCH-EER is obtained by maximizing w.r.t.\ the operating point, while minimizing w.r.t.\ the threshold. The minimization confines us to the ROCCH curve, while the maximization finds the most \emph{pessimistic} operating point on this curve. The ROCCH-EER therefore forms a tight upper bound on the Bayes error-rate that can be obtained with perfect calibration. By pushing down on the EER, we are pushing down the whole curve.

Another way to see this is the fact that $\minDCF(\effpi,1,1)$ is a \emph{concave} function (see figure~\ref{fig:rocchdet}). If we push down at the maximum of this curve (by trying to build a system that gets better EER) it cannot form a dent in the curve that violates concavity. If anything moves, the whole curve has to go down in such a way as to respect concavity.

This does \emph{not} guarantee that if we reduce ROCCH-EER, we will have reduced minDCF at \emph{all} operating points. Even if the value of the maximum is reduced, its position, $\effpi$, can move in such a way that error-rates can increase somewhere far from the maximum. This lateral movement is roughly analogous to tilting of the DET-curve. If, however, we want to target a specific region of operating points of interest, we can generalize this idea. This is shown in the next subsection.

\subsubsection{UER: Unequal-error-rate}
\def\UER{\text{UER}}
\def\Qmiss{\check P_\text{miss}}
\def\Qfa{\check P_\text{fa}}
We can generalize ROCCH-EER by considering a point, $[\Qfa,\Qmiss]$, on the ROCCH curve where $\Qfa=r\Qmiss$. For any $r\ne1$, this is an \emph{unequal-error-rate}. 

Such points also have the interpretation that they form tight upper bounds on minDCF. To see this, choose any costs such that: $\Cmiss=r\Cfa$. We can show~\cite{phd:Brummer} that there exists a point, $[\Qfa(r),\Qmiss(r)]$, on the ROCCH curve, such that:
\begin{align}
\label{eq:genEER}
\Cmiss \Qmiss(r) &= \Cfa \Qfa(r) = \max_{\pi} \minDCF(\pi,\Cmiss,\Cfa)
\end{align} 
The point on the curve depends just on the ratio $r$. By varying $r$ between zero and infinity, we can map out the whole ROCCH curve.\footnote{Interestingly, if we exchange max and min, the error-rates that satisfy $\min_\gamma \max_\pi \text{DCF}(\gamma|\pi,r,1)$, map out the steppy ROC as we vary $r$.} If we arbitrarily set $\Cfa=1$ and $\Cmiss=r$, we can define the unequal-error-rate as:
\begin{align}
\UER(r) &=\Qfa(r)=r\Qmiss(r) = \max_{\pi} \minDCF(\pi,r,1)
\end{align} 
Again, this value forms a tight upper bound of a concave function of $\pi$, so that using UER as optimization objective pushes down the whole curve. If we choose $r=\effpi$, then we will be targeting operating points in the vicinity of $\effpi$.

In summary, the whole ROC/DET curve has this `stiffness' property induced by the concavity, so that trying to optimize some point on the curve will tend to also improve the decision-making ability of the curve over a larger region.

\subsubsection{PRBEP}
Finally, we mention another variant on this idea, where we re-weight the error-rates to represent absolute error counts. By choosing $\Cmiss=T$, the number of target trials; and $\Cfa=N$, the number of non-target trials, the toolkit provides the functionality to compute the \emph{precision-recall-break-even-point}:
\begin{align}
\text{PRBEP} &= N\times\UER\bigl(\frac{T}{N}\bigr) = T\Qmiss=N\Qfa = \max_{\pi} \minDCF(\pi,T,N)
\end{align}
which represents the point on the ROCCH curve where the \emph{absolute} number of misses and false alarms are equal.\footnote{Since the ROCCH curve is an interpolation, this will in general not be a whole number.}

If the error-rates of the recognizer are low relative to the number of available evaluation trials, then this forms a sensible evaluation objective, which balances the two error-counts, keeping them both from becoming too small for as long as possible.

Here we prefer to present the result as an absolute number of errors, rather than as an error-rate, so that if the number of errors becomes small, the user is effectively warned that this is happening. 

PRBEP cannot be used for meaningful comparisons across databases of different sizes. It is meant for comparison of different systems on the same database.

\subsection{Fusion and Calibration}
\label{subsec:fusion}
\def\qvec{\mathbf{q}}
\def\rvec{\mathbf{r}}
\def\Wmat{\mathbf{W}}
The toolkit provides two solutions for calibration, which is the task of finding a mapping $\ell$, that maps scores to log-likelihood-ratios. In both cases, the mapping is `trained' on a supervised calibration database. One solution is non-parametric, based on isotonic regression. The other is parametric, based on logistic regression. The logistic regression solution generalizes also to a fusion recipe.\footnote{It is shown in~\cite{art:provost_mach_learn_2001}, that isotonic regression can also be used for fusion, but this is not yet implemented in the toolkit.}

The non-parametric calibration finds a solution that is (on the \emph{training} data) simultaneously optimal for \emph{any} sensible objective function\footnote{This is, any strict, or non-strict proper scoring rule, or Bayes risk criterion.} for measuring the goodness of calibration~\cite[Appendix~C]{phd:Brummer}. In practice however, we have found that the parametric solution usually performs better on independent test data.

\subsubsection{PAV: Non-parametric calibration}
The convention that the larger the score, the more it favours the target hypothesis, suggests that the calibration mapping, $\ell$, should be monotonically rising (isotonic)~\cite{inproc:Zadrozny_PAV_2002}. Since we have a finite number of training scores, each of which must be mapped to a log-likelihood-ratio, this can be done in a non-parametric way. We can independently choose the value for each point, subject only to the monotonicity constraint. This problem is known as \emph{isotonic regression} and an efficient implementation is given by the PAV (pool adjacent violators) algorithm, which we discuss in the next section.  

Attractive features of this solution are:
\begin{itemize}
	\item On training data, as mentioned above, it is optimal, no matter how you measure optimality.
	\item It corresponds exactly to $\minE$ (minDCF): If a data set is optimized with PAV, and then evaluated on the same data set with $\calE$ (DCF), then DCF = minDCF. 
	\item It also corresponds exactly to using the \emph{slope} of the ROCCH curve as calibrated likelihood-ratio~\cite{art:fawcett_mach_learn_2007}.
	\item The type of the score distribution is unimportant. In fact, the procedure is invariant to any monotonic warping of the scores. In contrast, the parametric logistic regression calibration solution below works best for approximately normal score distributions.
\end{itemize}

\subsubsection{Logistic regression: parametric fusion and calibration}
The toolkit provides a logistic regression solution, which can:
\begin{itemize}
	\item train a calibration mapping, $\ell(s)$, for a single system;
	\item train combination weights to fuse multiple subsystems into a single sub-system which outputs well-calibrated log-likelihood-ratios; and
	\item also incorporate certain kinds of side-information, or quality measures. 
\end{itemize}
All of this functionality is provided by optimization of the parameters of the following mapping:
\begin{align}
\ell_t &= a + \sum_{i=1}^N b_i s_{it} +  \qvec_t'\Wmat\rvec_t
\end{align} 
where $\ell_t$ is the fused and calibrated output log-likelihood-ratio for trial $t$; $N$ is the number of subsystems to be fused (if $N=1$, then the result is just calibration); $s_{it}$ is the score of subsystem $i$ for trial $t$; $\qvec_t$ and $\rvec_t$ are optional `quality vectors', derived from the two sides (enrol, verify) of trial $t$. The parameters to be optimized are the scalar offset $a$, the scalar combination weights $b_i$ and a symmetric matrix $\Wmat$, which effectively combines the two quality vectors into a quality score for the trial.

The parameters are optimized with logistic regression, which minimizes an objective function, which is very similar to the above-defined $\Cllr$. This objective function is the evaluation criterion for a \emph{supervised calibration database}, which must be provided by the user. Since the objective function is calibration sensitive, optimizing it causes the fused output to be well calibrated. See~\cite{art:Brummer_Fusion_TASLP_2007}, or ~\cite[Chapter 8]{phd:Brummer} for more details.

\section{Algorithms}
This section describes the key algorithms that help the toolkit to efficiently process very large sets of scores.

\subsection{Efficient DCF and minDCF}
This subsection describes efficient algorithms for computing DCF and minDCF. With more traditional implementations, computation of $\calE$ (DCF) and $\minE$ (minDCF), over the range required by a normalized Bayes error-rate plot, may take several minutes for large trial lists (a few million scores). By comparison, the implementation in the BOSARIS Toolkit takes a few seconds to execute.

\subsubsection{DCF}
To efficiently compute $\calE$, pool all the scores, $\calL=\ell_1,\ell_2,\ldots$, with all the different thresholds, $-\logit\effpi_i$, at which $\calE(\calL|\effpi_i)$ is to be evaluated. Sort them all together, in increasing order, keeping track of where the thresholds end up. The miss and false-alarm rates at threshold $i$ are given by
\begin{align}
\Pmiss(i) &= \{t_i - (D - i + 1)\} / T \\
\Pfa(i) &= \{N - (n_i - (D - i + 1))\} / N
\end{align}
where $t_i$ is the position of the $i$th threshold in the sorted list (after deleting non-target scores), $n_i$ is the position of the $i$th threshold in the sorted list (after deleting target scores), $D$ is the number of thresholds, $T$ is the number of target scores and $N$ is the number of non-target scores.  Equation~\ref{eq:def_ber} then gives $\calE$.

\subsubsection{minDCF}
To efficiently compute $\minE$, compute the vertices of the ROCCH curve, using the PAV algorithm (see section~\ref{sec:pav}). There are typically very few of these vertices and as shown in section~\ref{sec:rocch}, the original large ROC can be replaced with these vertices, without changing the value of minDCF\@.  Then use the last line of~\eqref{eq:minDCF}.

\subsection{The PAV Algorithm}
\label{sec:pav}
The PAV (pool adjacent violators) algorithm is central to the efficient implementation of many of the toolkit functions. We use it to efficiently compute the vertices\footnote{The vertices of the whole convex hull are the same as the vertices (cusps) of the piece-wise linear ROCCH curve.} of the ROCCH curve~\cite{art:fawcett_mach_learn_2007}. Once we have these vertices, we can compute minDCF, EER, UER, PRBEP and the non-parametric calibration mapping (see the relevant subsections in the theory section).

The PAV algorithm solves the problem of assigning a likelihood-ratio to each score in some supervised database of target and non-target scores. The likelihood-ratios are adjusted non-parametrically and independently, subject only to the monotonicity constraint that if the scores are sorted, then the likelihood-ratios must also be sorted. The PAV solution turns out to be simultaneously optimal for any proper scoring rule and therefore for any Bayes risk criterion, with any cost function and any prior~\cite[Appendix~C]{phd:Brummer}.

The PAV algorithm complexity is linear in the number of scores and the preceding sort has complexity of order $T\log(T)$. In our implementation, sorting and applying PAV takes a few seconds for a few million scores.  

\subsection{New logistic regression optimizer}
\label{sec:optimizer}
The BOSARIS Toolkit uses a general-purpose, unconstrained convex optimization algorithm to train the logistic regression fusion and calibration solutions. It uses a quasi-Newton method, which is faster, generally better behaved and converges to a better solution than the  conjugate gradient optimizer which was used in its predecessor, the FoCal Toolkit.\footnote{Available at: \url{http://sites.google.com/site/nikobrummer/focal}}

The new optimizer uses the \emph{trust region Newton conjugate gradient algorithm} for large-scale unconstrained minimization~\cite{book:Nocedal_Optimization_2006,art:Lin_TRNCG}.

\section{Code}
This section gives a high-level overview of some of the salient features of the implementation of the algorithms. More detail is available in the user manual which is distributed with the toolkit.

The current implementation is written in MATLAB, with an object-oriented API (application programmer's interface). The objects are not an essential part of the code,\footnote{MATLAB object oriented code does not scale well to large problems.} they are just a way to organize the API. If this type of interface turns out to be a hindrance rather than a help to users, it would be possible to replace this API.

The main feature of the code that remains to be highlighted in this last section is the efficient, binary, platform-independent score file format. The efficiency of the format relies on the assumption that trial lists can be represented as dense matrices, where the row and column indices are the two sides (enrol, verify) of a trial. We assume that each enrolment or each verification side is to be matched against many---or even all---others. (Such dense score matrices were necessary for ensuring an adequate number of non-target trials and therefore an adequate number of false-alarms at the new operating point, $\effpi=0.001$, of SRE'10.)

We use a platform-independent HDF5 binary score format to encourage interoperability with other tools. Text files would also give interoperability, but are much larger and much slower to process.

\subsection{Data}
\label{sec:data}

The code in the toolkit is primarily concerned with storing and manipulating the following data types:
\begin{description}
\item[indexes] list model and test segment names and indicate which pairs of model and test segment are in the trial list described by the index.   
\item[keys] are similar to indexes, but also give the answers i.e.\ which trials are target trials and which are non-target trials.
\item[scores] store scores for a list of trials (specified by an index or a key).  In addition to the actual scores, a score object contains all the information that an index describes.  
\item[quality measures] can be seen as scores for a model or test segment (instead of for a trial).  These can be fused with ordinary scores (see section~\ref{subsec:fusion}).  
\end{description}

Indexes can be used:
\begin{itemize}
\item for aligning scores from different systems before fusing them
\item for selecting parts of score objects of interest (e.g.\ those for male trials)
\item by external code that produces scores.  This code can load an index file which indicates which segment pairs to produce scores for.
\end{itemize}

Two score objects can be merged to make a new score object provided that they don't provide scores for the same trial.  Parts of score objects can be selected (to produce a new score object) either by using an index or by using lists of models or segments to discard.

\subsection{Plots}

The toolkit can produce two types of plots:
\begin{description}
\item[DET plots] (see section~\ref{subsec:roc}) either from points on the ROC or from the ROCCH curve.
\item[Normalized Bayes error-rate plots] (see section~\ref{sec:bayes_plots}).  Both minimum and actual Bayes error-rate curves can be plotted, as well as curves showing the contributions of the misses and false alarms, respectively, to those curves.  A vertical line indicating the operating point can be placed on the plot.
\end{description}

DR30 points (see section~\ref{subsec:sampling}) for misses and false alarms can be placed on both of types of plots.  

\subsection{Calibration}

The high level wrapper functions for calibration have two variants: those that train the calibration transformation on a single set~\footnote{By \emph{set}, we mean \emph{multiset}, because the collection should retain duplicate values.} of scores and then apply that transformation to the same set, and those that train the transformation on one set of scores (\emph{dev}) and apply it to another set (\emph{eval}).  A second partitioning of the functions can be made according to whether the transformation is affine or whether it uses the PAV algorithm (see section~\ref{sec:pav}).

\subsection{Fusion}

The main functions for doing fusion can again be divided (as for calibration) according to whether there is a set of unsupervised \emph{eval} scores in addition to the \emph{dev} scores or not.  There are separate wrapper functions for doing fusion when quality measures are to be used.

\subsection{Other functions}

There are functions for calculating EER, minimum DCF, actual DCF, PRBEP and the effective prior.

\subsection{File format}
\label{sec:files}

With approximately eight million trials in our development list for
SRE'10, loading and saving score files in text format became
unfeasible.  We therefore created a binary file format which both
reduced the size of the file on disk and made loading and saving
faster.  For example, one of our \emph{tel-tel} development files is about 60 times larger on disk in text format than in binary format and the binary file loads about 160 times faster than the text file.


The binary score files contain two lists and two matrices.  The lists
contain the model and test segment names.  One matrix contains the
scores as real numbers and the other matrix is a logical matrix of the
same size which indicates which scores correspond to valid trials.
The dimensions of the matrices are the number of models by the number
of test segments and the score at position $(i,j)$ is for a trial
between the $i$th model in the model list and the $j$th test segment
in the test segment list.

The toolkit provides both Matlab \emph{.mat} and HDF5 versions of the
binary format, as well as functions for converting between binary and
text formats.

\section{Acknowledgements}
We would like to thank our SRE'10 collaborators, BUT and CRIM, as well as the other participants of the BOSARIS Workshop for their contributions, especially Luk\'{a}\v{s} Burget, Old\v{r}ich Plchot and Nicolas {Scheffer}.

\bibliography{mybib}

\begin{thebibliography}{10}

\bibitem{phd:Brummer}
Niko {Br\"{u}mmer},
\newblock {\em Measuring, Refining and Calibrating Speaker and Language
  Information Extracted from Speech},
\newblock Ph.D. thesis, University of Stellenbosch, Stellenbosch, South Africa,
  Dec. 2010.

\bibitem{web:sre10}
The National~Institute of~Standards and Technology,
\newblock ``The {NIST} year 2010 speaker recognition evaluation plan,''
  \url{http://www.itl.nist.gov/iad/mig//tests/sre/2010/NIST_SRE10_evalplan.r6.pdf},
  Apr. 2010.

\bibitem{art:Brown_2001}
Lawrence~D. {Brown}, T.~Tony {Cai}, and Anirban {DasGupta},
\newblock ``Interval estimation for a binomial proportion,''
\newblock {\em Statistical Science}, vol. 16, no. 2, pp. 101--133, 2001.

\bibitem{inproc:DoddingtonsRule_98}
George~R. {Doddington},
\newblock ``Speaker recognition evaluation methodology: a review and
  perspective,''
\newblock in {\em Proceedings of {RLA2C} Workshop: Speaker Recognition and its
  Commercial and Forensic Applications}, Avignon, France, Apr. 1998, pp.
  60--66.

\bibitem{book:DeGroot_Opt_Stat_1970}
Morris~H. {DeGroot},
\newblock {\em Optimal Statistical Decisions},
\newblock McGraw-Hill, 1970.

\bibitem{inproc:brummer_odyssey04}
Niko {Br\"{u}mmer},
\newblock ``Application-independent evaluation of speaker detection,''
\newblock in {\em Proceedings of the Odyssey Speaker and Language Recognition
  Workshop}, Toledo, Spain, June 2004, pp. 33--40.

\bibitem{art:brummer_csl_2006}
Niko {Br\"{u}mmer} and Johan~A. {du Preez},
\newblock ``Application-independent evaluation of speaker detection,''
\newblock {\em Computer Speech and Language}, vol. 20, no. 2--3, pp. 230--275,
  2006.

\bibitem{incoll:VanLeeuwen_Eval_2007}
David~A. {van Leeuwen} and Niko {Br\"{u}mmer},
\newblock ``An introduction to application-independent evaluation of speaker
  recognition systems,''
\newblock in {\em Speaker Classification I: Fundamentals, Features, and
  Methods}, Christian M\"{u}ller, Ed., Lecture Notes in Computer Science /
  Lecture Notes in Artificial Intelligence, pp. 330--353. Springer, first
  edition, 2007.

\bibitem{art:DeGroot_Forecast_1983}
Morris~H. {DeGroot} and Stephen~E. {Fienberg},
\newblock ``The comparison and evaluation of forecasters,''
\newblock {\em The Statistician}, vol. 32, pp. 14--22, 1983.

\bibitem{book:Jaynes_PTTLOS_2003}
Edwin~T. {Jaynes},
\newblock {\em Probability Theory: The Logic of Science},
\newblock Cambridge University Press, 2003.

\bibitem{techrep:UCL_Dawid_139}
A.~Philip {Dawid},
\newblock ``Coherent measures of discrepancy, uncertainty and dependence, with
  applications to {B}ayesian predictive experimental design,''
\newblock Technical Report 139, Department of Statistical Science, University
  College London, Aug. 1998,
\newblock Online:
  \url{http://www.ucl.ac.uk/Stats/research/reports/abs94.html#139}.

\bibitem{phd:Ramos}
Daniel {Ramos-Castro},
\newblock {\em Forensic Evaluation of the Evidence Using Automatic Speaker
  Recognition Systems},
\newblock Ph.D. thesis, Universidad Aut\'{o}noma de Madrid, Madrid, Spain, Nov.
  2007.

\bibitem{art:Fawcett_PatRecLet_ROC_2006}
Tom {Fawcett},
\newblock ``An introduction to {ROC} analysis,''
\newblock {\em Pattern Recognition Letters}, vol. 27, no. 8, pp. 861--874, June
  2006.

\bibitem{inproc:Martin_EUROSPEECH_1997}
Alvin {Martin}, George {Doddington}, Terri {Kamm}, Mark {Ordowski}, and Mark
  {Przybocki},
\newblock ``The {DET} curve in assessment of detection task performance,''
\newblock in {\em Proceedings of the 5th European Conference on Speech
  Communication and Technology, {EUROSPEECH}}, Rhodes, Greece, Sept. 1997, pp.
  1895--1898.

\bibitem{art:provost_mach_learn_2001}
Foster~J. {Provost} and Tom {Fawcett},
\newblock ``Robust classification for imprecise environments,''
\newblock {\em Machine Learning}, vol. 42, no. 3, pp. 203--231, Mar. 2001.

\bibitem{inproc:Aversano_Evalita_2009}
Guido {Aversano}, Niko {Br\"{u}mmer}, and Mauro {Falcone},
\newblock ``{EVALITA} 2009 speaker identity verification ``application'' track
  organizer's report,''
\newblock in {\em Proceedings of {EVALITA}}, Reggio Emilia, Italy, Dec. 2009,
\newblock Online: \url{http://evalita.fbk.eu/proceedings.html}.

\bibitem{inproc:Zadrozny_PAV_2002}
Bianca {Zadrozny} and Charles {Elkan},
\newblock ``Transforming classifier scores into accurate multiclass probability
  estimates,''
\newblock in {\em Proceedings of the Eighth International Conference on
  Knowledge Discovery and Data Mining}, Edmonton, Alberta, {Canada}, 2002, pp.
  694--699.

\bibitem{art:fawcett_mach_learn_2007}
Tom {Fawcett} and Alexandru {Niculescu-Mizil},
\newblock ``{PAV} and the {ROC} convex hull,''
\newblock {\em Machine Learning}, vol. 68, no. 1, pp. 97--106, July 2007.

\bibitem{art:Brummer_Fusion_TASLP_2007}
Niko {Br\"{u}mmer}, Luk\'{a}\v{s} {Burget}, Jan~``Honza'' {\v{C}ernock\'{y}},
  Ond\v{r}ej {Glembek}, Franti\v{s}ek {Gr\'{e}zl}, Martin {Karafi\'{a}t},
  David~A. {van Leeuwen}, Pavel {Mat\v{e}jka}, Petr {Schwarz}, and Albert
  {Strasheim},
\newblock ``Fusion of heterogenous speaker recognition systems in the {STBU}
  submission for the {NIST} speaker recognition evaluation 2006,''
\newblock {\em {IEEE} Transactions on Audio, Speech, and Language Processing},
  vol. 15, no. 7, pp. 2072--2084, Sept. 2007.

\bibitem{book:Nocedal_Optimization_2006}
Jorge {Nocedal} and Stephen~J. {Wright},
\newblock {\em Numerical Optimization},
\newblock Springer, second edition, 2006.

\bibitem{art:Lin_TRNCG}
Chih-Jen {Lin}, Ruby~C. {Weng}, and S.~Sathiya {Keerthi},
\newblock ``Trust region {N}ewton method for large-scale logistic regression,''
\newblock {\em Journal of Machine Learning Research}, Sept. 2008.

\end{thebibliography}
\bibliographystyle{IEEEbib}
\end{document}